\documentclass[12pt, a4paper]{article}

\usepackage{lscape}
\usepackage{array}
\usepackage{booktabs}

\newcolumntype{L}[1]{>{\raggedright\let\newline\\\arraybackslash\hspace{0pt}}p{#1}}
\newcolumntype{C}[1]{>{\centering\let\newline\\\arraybackslash\hspace{0pt}}p{#1}}
\newcolumntype{R}[1]{>{\raggedleft\let\newline\\\arraybackslash\hspace{0pt}}p{#1}}

\def\bea{\begin{eqnarray*}}
\def\eea{\end{eqnarray*}}
\def\bean{\begin{eqnarray}}
\def\eean{\end{eqnarray}}
\def\bi{\begin{itemize}}
\def\ei{\end{itemize}}

\usepackage{amsmath}
\usepackage{amssymb}
\usepackage{amsfonts}
\usepackage{amsthm}
\usepackage{latexsym}
\usepackage{color}
\usepackage{epsfig}
\usepackage{hyperref}
\usepackage{longtable}
\usepackage{multirow}
\usepackage{lscape}
\usepackage[normalem]{ulem}   
\usepackage{ulem}
\usepackage[table]{xcolor}

\newcommand{\aje}{AJE }

\newcommand{\ajep}{AJE. }

\newcommand{\ndz}{{\color{white}0}}

\begin{document}

\title{Survival analysis for AdVerse events with VarYing follow-up times (SAVVY): summary of findings and a roadmap for the future of safety analyses in clinical trials}

\author{
    Kaspar Rufibach\thanks{Methods, Collaboration, and Outreach Group, Product Development Data Sciences (MCO), Hoffmann-La Roche Ltd, Basel, Switzerland}\\
    Jan Beyersmann\thanks{Institute of Statistics, Ulm University, Ulm, Germany}\\
    Tim Friede\thanks{Department of Medical Statistics, University Medical Center G\"ottingen} \\
    Claudia Schmoor\thanks{Clinical Trials Unit, Faculty of Medicine and Medical Center, University of Freiburg}\\
    Regina Stegherr\thanks{Charit\'{e} - Universit\"atsmedizin Berlin, Corporate Member of Freie Universit\"{a}t Berlin and Humboldt-Universit\"{a}t zu Berlin, Institute of Biometry and Clinical Epidemiology}
 }

\maketitle

\begin{abstract}
  The SAVVY project aims to improve the analyses of adverse events (AEs) in clinical trials through the use of survival techniques appropriately dealing with varying follow-up times and competing events (CEs). This paper summarizes key features and conclusions from the various SAVVY papers. Through theoretical investigations using simulations and in an empirical study including randomized clinical trials from several sponsor organisations, biases from ignoring varying follow-up times or CEs are investigated. The bias of commonly used estimators of the absolute and relative AE risk is quantified. Furthermore, we provide a cursory assessment of how pertinent guidelines for the analysis of safety data deal with the features of varying follow-up time and CEs. SAVVY finds that for both, avoiding bias and categorization of evidence with respect to treatment effect on AE risk into categories, the choice of the estimator is key and more important than features of the underlying data such as percentage of censoring, CEs, amount of follow-up, or value of the gold-standard. The choice of the estimator of the cumulative AE probability and the definition of CEs are crucial. SAVVY recommends using the Aalen-Johansen estimator (AJE) with an appropriate definition of CEs whenever the risk for AEs is to be quantified. There is an urgent need to improve the guidelines of reporting AEs so that incidence proportions or one minus Kaplan-Meier estimators are finally replaced by the \aje with appropriate definition of CEs. 
\end{abstract}


\textit{Running head:} SAVVY overview.

%
%

\section{Background}

In randomized clinical trials (RCT), an essential part of the benefit-risk assessment of treatments is the quantification of the risk of experiencing adverse events (AEs), and comparing these risk between treatment arms. Methods commonly employed to quantify absolute adverse event (AE) risk either do not account for varying follow-up times, censoring, or for competing events (CEs), although appreciation of these are important in risk quantification, see e.g. O'Neill\cite{oneill_87} and Procter and Schumacher\cite{proctor_16}.

Analyses of AE data in clinical trials can be improved through use of survival techniques that account for varying follow-up times, censoring and CEs. The \aje\cite{aalen_78, allignol_16} can be considered the non-parametric gold-standard method when quantifying absolute AE risk. The reason is that the \aje is the standard (non-parametric) estimator that accounts for CEs, censoring, and varying follow-up times simultaneously, and, being non-parametric, does not rely on restrictive parametric assumptions, such as constant hazards. Any other estimator of AE probability, such as incidence proportion, probability transform incidence density, or one minus Kaplan-Meier, delivers biased estimates in general.

To quantify that bias for all these methods in an ideal scenario Stegherr et al.\cite{stegherr_20} ran a comprehensive simulation study. Two key findings were (1) that ignoring CEs is more of a problem than falsely assuming constant hazards by the use of the incidence density and (2) that the choice of the AE probability estimator is crucial for the estimation of relative effects, i.e. comparison between groups.

To illustrate and further solidify these simulation-based results with real data the SAVVY consortium, a collaboration between nine pharmaceutical companies and three academic institutions, meta-analyzed data from seventeen randomized controlled trials (RCT).

In this review article we summarize the results of the empirical study, reported in two separate publications: Stegherr et al.\cite{onesample} was concerned with estimation of AE risk in one treatment group and Rufibach et al.\cite{rufibach_23} with the comparison of AE risks between two groups in an RCT. A cursory assessment how relevant guidelines recommend to estimate AE risk is given with a call for updates. We conclude with a discussion.


\section{Scientific questions of the SAVVY project}

The overarching scientific questions of SAVVY can be phrased as follows:
\begin{enumerate}
    \item[1)] For estimation of the probability of an AE, how biased are commonly used estimators, especially the incidence proportion and one minus Kaplan-Meier, in presence of censoring, varying follow-up between patients, CEs, and in the case of incidence densities a restrictive parametric model?
    \item[2)] What is the bias of common estimators that quantify the relative risk of experiencing an AE between two treatment arms in a RCT?
    \item[3)] Can trial characteristics be identified that help explain the bias in estimators?
    \item[4)] How does the use of potentially biased estimators impact qualification of AE probabilities and relative effects in regulatory settings?
\end{enumerate}

Within the SAVVY project, these questions were approached in two ways: First, in Stegherr et al.\cite{stegherr_20}, via simulation of clinical trial data. This approach has the advantage that the true underlying data generating mechanism is specified by the researcher and therefore known. This allows to exactly quantify the bias of a given estimator, i.e. to answer 1) and 2) above (it would also allow to answer Question 4, but that was not addressed in Stegherr et al.\cite{stegherr_20}). Second, in Stegherr et al.\cite{onesample} and Rufibach et al.\cite{rufibach_23}, biases of commonly used estimators of absolute and relative AE risks were estimated by comparing them to the best available estimator. Having real clinical trial data available also allows to answer Question 3 above, through meta-analytic methods.

\subsection{Competing events and their connection to the ICH E9 estimands addendum}
\label{sec:CE}

In what follows we will use {\it competing event} (apart from direct quotes) and consider it synonymous to {\it competing risk}. 

An important but largely unrecognized aspect when quantifying AE risk is the likely presence of CEs. Gooley et al.\cite{gooley_99} define a CE as 
\begin{quote}
``We shall define a competing risk as an event whose occurrence either precludes the occurrence of another event under examination or fundamentally alters the probability of occurrence of this other event.''
\end{quote}
whereas the ICH E9(R1) estimands addendum\cite{iche9r1} defines an intercurrent event as 
\begin{quote}
``Events occurring after treatment initiation that affect either the interpretation or the existence of the measurements associated with the clinical question of interest.''
\end{quote}
These two definitions appear to be, if not the same, then at least very related. However, the ICH E9(R1) addendum does not discuss CEs, so it is not entirely clear how to embed CEs into the addendum framework, i.e. whether and if yes which of the proposed strategies of the estimand addendum applies to the CE situation. More research and discussion is needed to align CEs (if necessary at all), and the analysis of  complex time-to-event data with the addendum. 

Stegherr et al.\cite{onesample} and Rufibach et al.\cite{rufibach_23} took a pragmatic approach and defined events as ``competing'' that preclude the occurrence or recording of the AE under consideration in a time-to-first-event analysis. Specifically, one important CE is death before AE. In addition, any event that would both be viewed from a patient perspective as an event of his/her course of disease or treatment and would stop the recording of the interesting AE was viewed as a CE. Since all these CEs apart from {\it death} may be prone to some subjectivity in the empirical analysis reported in Stegherr et al.\cite{onesample} and Rufibach et al.\cite{rufibach_23} a variant of the estimators with a CE of {\it death only} was also considered. Since results were in line with the broader definition of CEs as given above we omit the results for the {\it death only} variant here.

\subsection{Estimation methods}

A precise mathematical definition of all estimators of the probability of an AE that were compared in SAVVY is provided in Stegherr et al.\cite{stegherr2019survival}, a prospectively published statistical analysis plan for the SAVVY project. A short introduction in all estimators is also provided in Stegherr et al.\cite{onesample}. In this overview article we only provide very brief descriptions of the considered estimators. By far the most commonly used estimator, e.g. in standard safety reporting that enters benefit-risk assessment for the approval of new medicines, to estimate the risk of an adverse event up to a maximal observation timepoint $\tau$ is the {\it incidence proportion}\cite{allignol_16}. It simply divides the number of patients with an observed AE on $[0, \tau]$ in group $A$ by the number of patients in group $A$. The incidence proportion is an estimator of the probability that an AE happens in the interval $[0, \tau]$, {\it and} that this AE is observed, i.e. not censored. This illustrates that the incidence proportion is not properly dealing with censored observations. However, it correctly accounts for CEs, see Allignol et al.\cite{allignol_16} for an exemplary illustration of that feature. To account for the differing follow-up times between patients researchers and guidelines (see Table~\ref{tab:guidelines}) often advocate to use the {\it incidence density}, where the number of AEs in the nominator is divided by the total patient-time at risk instead of simply the number of patients. As such, the incidence density does not directly estimate the probability of an AE, but rather the AE hazard. As described in Stegherr et al.\cite{stegherr2019survival}, this hazard estimator can easily be transformed to indeed estimate the probability of an AE. However, it only does so assuming that the AE hazard is constant, i.e. the probability-transformed incidence density is a fully parametric estimator. In addition, as such it does not correctly account for CEs, but it can be modified to do so, leading to the {\it probability transform incidence density accounting for CEs}. Researchers are often aware of the inability of the incidence proportion to properly deal with varying follow-up times and censoring. As a remedy they (and many guidelines, see Table~\ref{tab:guidelines}) then advocate to consider {\it time to AE} and estimate the probability of an AE by reading off the one minus Kaplan-Meier estimator at a timepoint of interest, e.g. the end of observation time $\tau$ or at an earlier timepoint. While indeed, this estimator properly accounts for varying follow-up times and censoring the question remains how to deal with CEs. Numerous papers have been written\cite{gooley_99, allignol_16} and providing technical arguments explaining why the Kaplan–Meier estimator is a biased estimator of the probability of an AE. Intuitively, one minus the Kaplan–Meier estimator estimates the distribution function of the time of interest, that is, it tends towards one as we move to the right on the time axis. However, if we add up the probability of an AE and the probability of a CE in a truly CE scenario that also must be equal to one, implying that the probability of an AE is {\it strictly smaller} than one. As a consequence, the Kaplan–Meier estimator to estimate the probability of an AE will be biased upwards. 
Finally, there is an estimator that at the same time accounts for (random or independent) censoring, respects varying follow-up times, accounts for CEs in the right way, is fully nonparametric and therefore free of bias introduced by any of these processes: the \aje\cite{aalen_78}. It is therefore considered the gold standard estimator. In the empirical analysis of the 17 RCTs in the SAVVY project, it served as a benchmark against which all estimators were measured against. The term {\it bias} was therefore used for deviations of the estimators from this benchmark estimator, or gold standard. For an evaluation of the true bias, i.e. the deviation of estimators to the true underlying value, we refer to Stegherr et al.\cite{stegherr_20}.

Table~1 in Stegherr et al.\cite{onesample} concisely summarizes the properties of each considered estimator with respect to whether it accounts for censoring, CEs, and whether it makes a parametric assumption, and we therefore reproduce it here in Table~\ref{tab:estimatorsbias} for the estimators discussed here.

\begin{table}[h!]
\small\sf\centering
\caption{Overview whether the estimators deal with the possible sources of bias, reproduced from Stegherr et al.\cite{onesample}.
\label{tab:estimatorsbias}}
 \begin{tabular}{lccc}
 \toprule
&Accounts for & Makes no constant & Accounts for \\
&censoring & hazard assumption & CEs\\
\midrule
Incidence proportion & No & Yes & Yes\\
1-Kaplan-Meier & Yes & Yes & No\\
gold-standard \aje & Yes & Yes & Yes \\
\bottomrule
\end{tabular}
\end{table}

\subsection{Quantification of bias - the SAVVY project}

One of the goals of the SAVVY project is to quantify the bias of standard estimators of the probability of an AE. Based on simulations, i.e. comparing estimated to true underlying values from which the data was simulated, this has been done in Stegherr et al.\cite{stegherr_20}. Key findings in this study were that ignoring CEs is more of a problem than falsely assuming constant hazards. The one minus Kaplan-Meier estimator may overestimate the true AE probability by a factor of four at the largest observation time. Moreover, the choice of the AE probability estimator is crucial for group comparisons.

To confirm these results in real clinical trial data three academic institutions and nine pharmaceutical companies collaborated within the SAVVY consortium. In order not to have to share the trial data macros to compute the above estimators were developed centrally. Companies then ran these macros on trial data of their choice and only returned high-level results (estimated values of AE probabilities for each estimator and some basic trial characteristics) to the central data analysis unit hosted at one of the academic institutions. The central data analysis unit then meta-analyzed these results to quantify biases for estimators and impact on regulatory decision-making. A statistical analysis plan for these analyses was published\cite{stegherr2019survival}. Results for estimation of the probability of an AE in one arm are reported in Stegherr et al.\cite{onesample} and for the comparison between two groups in Rufibach et al.\cite{rufibach_23}.

In this overview paper we focus attention on summarizing results for the most commonly used estimators of AE risks, namely the incidence proportion and one minus Kaplan-Meier in comparison to the gold standard AJE. Furthermore, we report results for the maximal evaluation time as defined in the above two papers, i.e. the latest time at which a dataset has an observation, either event or censored. Results for shorter evaluation times were in line with those for the maximal evaluation time.

\section{Results}

Ten organisations provided 17 trials including 186 types of AEs (median $8$; interquartile range $[3, 9]$). Twelve (71.6\% out of 17) trials were from oncology, nine (52.9\%) were actively controlled and eight (47.1\%) were placebo controlled. The trials included
between 200 and 7171 patients (median: 443; interquartile range $[411, 1134]$). 

For one of the 17 trials, details of the trial and the analysis of one of the AEs by the different methods investigated in this paper are presented in Stegherr et al.\cite{stegherr_20}.

Note that for RRs only those AEs were considered where neither the estimated AE probability in the experimental arm nor the estimated AE probability in the control arm is 0 ($n = 156$ for incidence proportion and $n = 155$ for one minus Kaplan-Meier). This also applies to the categories in Table~\ref{tab:ipkm}.

\subsection{Empirical bias of common estimators of absolute risk with respect to the gold-standard AJE}

For the comparison of the AE probabilities Stegherr et al.\cite{onesample} focussed on the experimental arm. Median follow-up was 927 days (interquartile range $[449, 1380]$). The median of the gold-standard \aje was 0.092 (minimum 0 and maximum 0.961), i.e. the estimated probability of an AE was 9.2\% on average over all 186 AEs over all trials. The one minus Kaplan-Meier estimator was on average about 1.2-fold larger than the \aje and the probability transform of the incidence density ignoring CEs was even two-fold larger. The average empirical bias (i.e. the difference between the considered estimator and the gold-standard AJE) using the incidence proportion was less than 5\%. Assuming constant hazards using incidence densities was hardly an issue provided that CEs were accounted for. However, beyond these average biases it was striking how extreme the bias could become: For one minus Kaplan-Meier, in our empirical analysis using real clinical trial data, we observed an overestimation of the AE probability up to a factor of five, whereas for the incidence proportion we observed underestimations of up to a factor of three. This is in line with the findings of the simulation study in Stegherr et al.\cite{stegherr_20} and illustrates that using too simplistic estimators for the probability of an AE can be truly misleading. To evaluate what study characteristics impact the bias a meta-regression was performed. For that, the response variable was defined as the AE probability estimates obtained with the different estimators divided by the AE probability obtained with the gold-standard AJE, considered on the log-scale. Covariates were the proportion of censoring, the evaluation time point $\tau$ (i.e., the maximal time to event in years -- AE, CE or censoring -- observed), and the size of the AE probability estimated by the gold-standard \ajep The meta-regression showed that the bias depended primarily on the amount of censoring and on the amount of CEs.

Finally, according to the European Commission's guideline on summary of product characteristics (SmPC)\cite{smpc} and based on the recommendations of the {\it Council for International Organizations of Medical Science} (CIOMS) Working Groups III and V\cite{AEcat}, the AE risk is classified into frequency categories which are defined by `very rare', `rare', `uncommon', `common', and `very common' when the risk is $<$ 0.01\%,$<$ 0.1\%,$<$ 1\%,$<$ 10\%,$\ge$ 10\%, respectively. Stegherr et al.\cite{onesample} (Table 2) assigned these categories to AE probability estimates from all estimators (and all 186 AEs) and compared them to the categories resulting from the AJE. As an example, systematic overestimation of the one minus Kaplan-Meier resulted in upgrading of 2/6 AEs from `uncommon' to `common' and 14/86 from `common' to `very common'.

\subsection{Empirical bias of probability-based estimators of relative risk with respect to the gold-standard AJE}

Naively, one could ask the question whether when we have biased estimates of the AE probability in two groups, that go in the same direction in both groups, and want to compare them, the bias ``cancels out''. To assess this hypothesis Rufibach et al.\cite{rufibach_23} extended the work from Stegherr et al.\cite{onesample} to quantify the bias when comparing AE risks between groups in a randomized trial. The focus of Rufibach et al.\cite{rufibach_23} is not to define what a fit-for purpose estimand to quantify safety risk could be, but rather to evaluate statistical properties of commonly used estimators in the presence of varying follow-up and CEs. A thorough discussion of effect measures to quantify the relative risk is given there as well. Without going into causal or estimand details, the effects to be compared between groups are to be understood as total effects, comparing patients' AE risk in this world and in the presence of CEs. The estimators that were finally assessed are 
\begin{itemize}
\item the risk ratio (RR) $\widehat{RR} = \hat q_E / \hat q_C$ for estimators $\hat q_E$ and $\hat q_C$ of AE probabilities calculated at a specific evaluation time within each treatment arm (E for experimental, C for control),
\item the hazard ratio (HR) for both, the AE of interest and the CE.
\end{itemize}
In the one-sample case, estimates of AE probabilities were benchmarked on the gold-standard \ajep This, because the latter is a fully nonparametric estimator that accounts for censoring, does not rely on a constant hazard assumption, and accounts for CEs, as discussed above. So, as a straightforward extension for the comparison of AE probabilities between two arms using the RR, we benchmark the latter on the RR estimated using the \aje in each arm, with variance derived using the delta rule. The gold-standard for estimates of the HR will be the HR from Cox regression. This, because the latter is typically used to quantify a treatment effect not only for efficacy, but also for time-to-first-AE type endpoints. Variances of comparisons of different estimators of the RR and HR will be received via bootstrapping, because of the dependency of estimators computed on the same dataset, see Stegherr et al.\cite{stegherr_20}.

\begin{figure*}[ht]
\setkeys{Gin}{width=1\textwidth}
\begin{center}
\caption{Ratio of RRs estimated with estimator of interest and the gold-standard \ajep}
\label{fig:rr}
\includegraphics{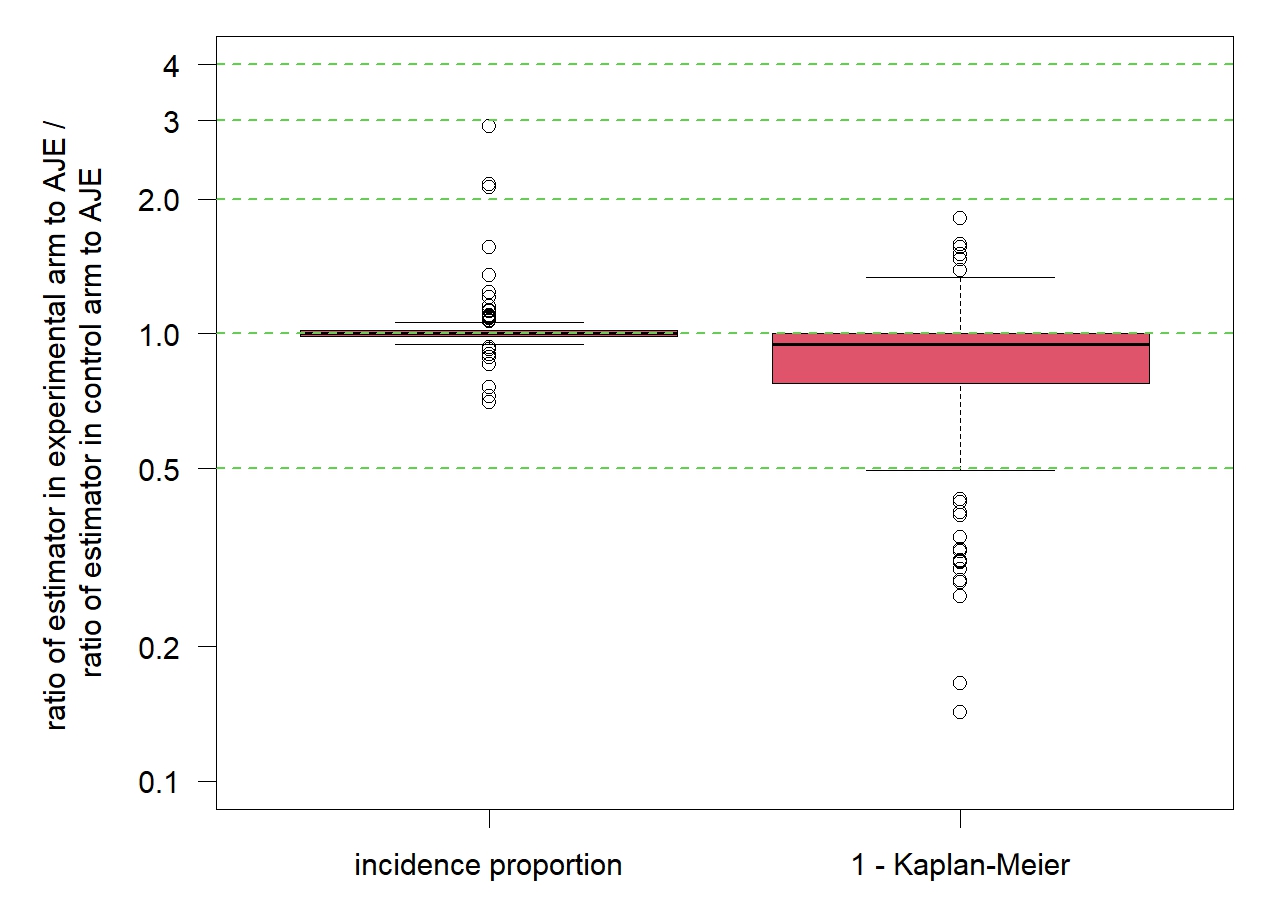}
\end{center}
\end{figure*}

For the one-sample case of estimation of absolute AE risk direction of biases can be derived analytically: incidence proportion systematically under- and one minus Kaplan-Meier systematically overestimates the true AE risk. However, there is no such derivation possible for direction of the bias for the RR or HR. Rufibach et al.\cite{rufibach_23} found the RR based on the incidence proportion overestimates the RR computed using the gold-standard \aje by up to a factor of three, and one minus Kaplan-Meier underestimates up to a factor of four, see Figure~\ref{fig:rr}. Interestingly, dividing the two biased estimates of the AE probability based on the incidence proportion, which both tend to {\it underestimate} the true AE probability, leads to an estimate of the RR that on average performs comparably to the \ajep Apart from shedding light on estimation quality of the incidence proportion and one minus Kaplan-Meier to estimate the RR we conclude that different patterns of under- or overestimation of absolute AE probabilities can lead to similar performance for RR. This implies that in general, one cannot conclude how an estimator of the relative AE risk performs based on looking how these same estimators performs on estimation of arm-wise AE probabilities.

As discussed in Stegherr et al.\cite{onesample} one reason for the good performance of the incidence proportion might be a high amount of CEs before possible censoring. However, not only the proportion of censoring but also the timing of the censoring are relevant.

Meta-analyses confirmed the above impressions. Meta-regressions showed that (1) the key difference between estimators lies in the value of the average RR and (2) the impact of covariates is overall limited, compared to the average RR the estimated coefficients are close to 1. This emphasizes that the choice of the estimator is key, and that this holds true over a wide range of possible data configurations quantified through the considered covariates.

A key finding of Rufibach et al.\cite{rufibach_23} was that categorization of evidence based on RR crucially depends on the estimator one uses to estimate the RR. We therefore reproduce Table~1 in their paper here in a version trimmed down to incidence proprtion and one minus Kaplan-Meier, see Table~\ref{tab:ipkm}. Overall, we find quite a number of switches to neighboring categories, more so than for estimators of the absolute AE risk in one arm. Reasons for switches are wider CIs of the \aje as well as RR estimates / CI bounds that are close to the cutoffs between categories. As the incidence proportion on average estimates the RR well, we see a similar number of switches to a higher ($n = 8$, below the diagonal in Table~\ref{tab:ipkm}) and lower ($n = 9$) evidence category. Not surprisingly, for one minus Kaplan-Meier that underestimates the RR with respect to the gold-standard \aje we see relevantly more switches to a lower than higher evidence category, namely $n = 41 / n = 16$, $32/8$, and $28/6$, respectively.

In summary, the choice of the estimator of the RR does have a relevant impact on the conclusions.
Note that there is no universally accepted standard how one should combine a point estimate and its associated variability, in our case RR, into evidence categories. As an example, Rufibach et al.\cite{rufibach_23} used a categorization motivated by the methods put forward by the IQWiG\cite{iqwigmethodenpaper} for severe AEs (Table 14) to be used for the German benefit-risk assessment. 

\begin{table}
\centering
\caption{The impact of the choice of relative effect estimators, incidence proprtion and one minus Kaplan-Meier, for AE probabilities on qualitative conclusions. Diagonal entries are set in bold face. Deviations from the gold-standard \aje are the off-diagonal entries. Off-diagonal zeros are omitted from the display.}
\label{tab:ipkm}
 \begin{tabular}{@{}c@{}c@{}c@{}c@{}c@{}c@{}clcccc}
 \toprule
&&& &&&&& \multicolumn{4}{c}{gold-standard Aalen-Johansen}\\
&&&&&&& &(0) no effect & (a) minor & (b) considerable & (c) major \\
\midrule
&&&     \multirow{4}*{\rotatebox{90}{incidence}} & \multirow{4}*{\rotatebox{90}{proportion}}
&&&     (0) no effect    &\textbf{84} & 5           & \ndz        & \ndz        \\
&&&&&&& (a) minor        & 3          & \textbf{10} & 2           & \ndz        \\
&&&&&&& (b) considerable & 1          & 2           & \textbf{12} & 2           \\
&&&&&&& (c) major        & 1          & \ndz        & 1           & \textbf{33} \\
\midrule
&\multirow{5}*{\rotatebox{90}{one}}&\multirow{5}*{\rotatebox{90}{minus}}&     \multirow{5}*{\rotatebox{90}{ Kaplan-}}& \multirow{5}*{\rotatebox{90}{Meier}}
&&&     (0) no effect    &\textbf{84} & 9           & 4           & 8           \\
&&&&&&& (a) minor        & 3          & \textbf{6}  & 3           & 3           \\
&&&&&&& (b) considerable & 2          & 1           & \textbf{7}  & 5           \\
&&&&&&& (c) major        & \ndz       & 1           & 1           & \textbf{18} \\
\bottomrule
\end{tabular}
\end{table}

\subsection{Empirical bias of hazard-based estimators of relative risk with respect to the gold-standard AJE}

As for hazard-based inference it is important to note that, even if the event of interest is AE, effects on all other CEs is generally recommended for any (hazard-based) analysis of CEs\cite{latouche_13}. Rufibach et al.\cite{rufibach_23} found that the hazard of AE is generally larger for the experimental compared to the control arm, meaning that the instantaneous risk of AE is typically higher in the experimental arm of an RCT, not unexpectedly. For the hazards of CEs, for both types, what we find is that the hazard in the experimental arm is generally lower than in the control arm, i.e. there is an effect of the experimental treatment on the CE. If we simply censored at CEs we would thus introduce arm-dependent censoring, a feature that may lead to biased effect estimates\cite{schemper_96, clark_02}. The ratio of the incidence densities of the AE in the two arms underestimates with respect to the Cox regression HR while for the other two endpoints on the median they turn out to be approximately unbiased compared to the Cox HR, with a tendency to overestimation when accounting for all CEs. To appreciate the differences between the two estimators of the RR based on hazards, i.e. the incidence density ratio and the gold-standard Cox regression HR, recall the properties of the two methods: Both properly account for censoring and they properly estimate event-specific hazards, or rather the relative effect based on these. The only difference between the two methods is what they assume about the shape of the underlying hazard: the incidence density assumes them to be {\it constant} up to the considered follow-up time, which also implies that they are {\it proportional}. 

The impact of the use of the different estimators on the conclusions derived from the comparison of treatment arms was again investigated by the use of categories. These are typically derived from comparing the confidence interval (CI) of the RR to thresholds. In contrast to the usual IQWiG procedure, however, they did not only categorize the benefit of a therapy, but also the harm, where they did not distinguish between a positive and a negative treatment effect. Four categories were possible: (0) ``no effect'' if 1 is included in the CI, (a) ``minor'' (``gering'') if the upper bound of the CI is in the interval $[0.9, 1)$ for a RR$<1$ or the lower bound in the interval $(1,1.11]$ for a RR$>1$, (b) ``considerable'' (``betr\"achtlich'') if the upper bound of the CI is in the interval $[0.75,0.9)$ for a RR$<1$ or the lower bound in the interval $(1.11,1.33]$ for a RR$>1$, and (c) ``major'' (``erheblich'') if the upper bound is smaller than 0.75 for a RR$<1$ or the lower bound greater than 1.33 for a RR$>1$. The same categorization was used for the HR instead of RR.

Rufibach et al.\cite{rufibach_23} have considered two effect measures to quantify the RR of an AE in two arms: the RR based on AE probabilities and the HR computed from Cox regression. Rufibach et al.\cite{rufibach_23}'s analyses revealed that all the considered estimators are overall inferior to the two gold-standards that were considered, either the RR based on the arm-wise \aje or the HR based on Cox regression. A question that remains is whether the qualitative conclusions drawn based on the two gold-standards are relevantly different when relying on the criteria put forward by the IQWiG (Table 14 in their general methods document\cite{iqwigmethodenpaper}), see Table~9 in Rufibach et al.\cite{rufibach_23}, which we reproduce here as Table~\ref{tab:RRcomp}. Quite some different classifications based on the two estimators of the RR were observed. However, this is not a surprise, as the {\it estimand} the two methods look at is not the same (see Varadhan et al.\cite{varadhan_10} for an exposition of this issue): Cox HR quantifies a relative effect based on an endpoint of AE \emph{hazard}, whereas RR based on gold-standard Aalen-Johansen is based on a comparison of \emph{probabilities} at a evaluation time, see Rufibach et al.\cite{rufibach_23} for an extended discussion of this issue. 

\begin{table}[h]
\centering
\caption{Conclusions of the RR calculated with the gold-standard Aalen-Johansen estimator compared to the conclusions of the HR calculated with the Cox model. The table shows the analysis of those $n=94$ AEs that were observed with an absolute frequency of $\ge 10$ in each arm. Off-diagonal zeros are omitted from the display.}
\label{tab:RRcomp}
\begin{tabular}{llcccc}
\toprule
                                                              &                  & \multicolumn{4}{@{}c@{}}{{HR Cox for AE}}              \\
                                                              &                  &(0) no effect & (a) minor  & (b) considerable & (c) major   \\
\midrule
\multirow{4}{*}{\shortstack{RR gold-standard\\ Aalen-Johansen}} & (0) no effect    & \textbf{42}  & 3          & 3                & 1           \\
                                                              & (a) minor        & 9            & \textbf{2} & 1                & \ndz        \\
                                                              & (b) considerable & 4            & 1          & \textbf{3}       & 2           \\
                                                              & (c) major        & 2            & \ndz       & 4                & \textbf{17} \\
\bottomrule
\end{tabular}
\end{table}

\subsection{Overall conclusions from the empirical analysis}

To conclude this section, based on theoretical and empirical evidence, Stegherr et al.\cite{onesample} clearly recommend the \aje as the non-parametric, unbiased estimator in the presence of both CEs and censoring. If a parametric analysis based on incidence densities is considered, they strongly recommend to incorporate incidence densities of CEs as they detail in their paper. This is also preferable over the one minus Kaplan-Meier approach. A conclusion of the empirical study of the SAVVY project not discussed here was also that ignoring CEs appeared to be worse than assuming constant hazards. This illustrates the importance of careful consideration of CEs when aiming at properly estimating and comparing AE risks.

\section{Analysis of safety in key guidelines}

To understand the landscape of existing guidelines on safety reporting which, ideally, at some point will be updated based on the conclusions from SAVVY, we reviewed an opportunistic sample of these and collected results in Table~\ref{tab:guidelines}. Many of these guidelines mention at least that varying follow-up times may be relevant to quantify AE risk. For example, the CIOMS Working Group VI handbook\cite{cioms_05}, which forms the basis of several guidances, already admits that ``ICH Guideline E3 mentions survival analysis methods for analysing safety data, but it appears that this has often not been followed in practice.'' Looking at Table~\ref{tab:guidelines} two aspects are remarkable: First, the heterogeneity in how related guidelines treat the issues of varying follow-up, use of the incidence density, constant hazard assumption for the latter, proposing life-table or one minus Kaplan-Meier techniques. Second, the complete absence of any consideration of CEs, although at least death seems to be quite an obvious CE in estimation of AE risk, with many others potentially relevant. Taken together with the frequent mentioning of life-table / one minus Kaplan-Meier methods to account for varying follow-up time this is specifically concerning given the findings in Stegherr et al.\cite{onesample} and Rufibach et al.\cite{rufibach_23} about the bias of one minus Kaplan-Meier in presence of CEs.

Overall, it appears somewhat suprising that all guidelines exhibit awareness of the varying follow-up time issue and even discuss potential remedies, but that in routine reporting of safety and quantification of AE risks the incidence proportion appears still to be the overwhelmingly preferred approach.

\begin{landscape}
\begin{table}[htbp]
\centering
\caption{Coverage of relevant time-to-event aspects for quantification of AE risk for an opportunistic sample of safety guidelines. "x" means that this aspect is mentioned in the respective guideline.}
\label{tab:guidelines}
\begin{tabular}{|C{5cm}|C{3cm}|C{3cm}|C{4.5cm}|C{3cm}|C{3cm}|} \hline
{\bf Guideline} &{\bf Acknowledges\newline varying FU} & {\bf Proposes \newline incidence density} & {\bf Acknowledges constant \newline hazard assumption} & {\bf Proposes life-table / \newline one minus Kaplan-Meier}& {\bf Acknowledges \newline CEs}\\ \hline
ICH E3\cite{iche3} &x &  && x & \\ \hline
ICH E9\cite{iche9} &x & x & & x & \\ \hline
SmPC\cite{smpc} &x & x &  & & \\ \hline
CIOMS\cite{cioms_05} &x & x &x  & x& \\ \hline
FDA premarketing\cite{FDA_05_premarket} &x & x &x  & x& \\ \hline
CONSORT Harm 2022 update\cite{junqueira_23} &x &  &  & & \\ \hline
\end{tabular}
\end{table}
\end{landscape}

\section{Implementation}

All methods introduced in the SAVVY project have been implemented in the R package savvyr\cite{savvyr}. This package is available from CRAN.

\section{Discussion}

\subsection{Main conclusion of the SAVVY project}

The main conclusion from the SAVVY project is that commonly used methods such as incidence proportions, incidence densities (with and without ignoring CEs), and one minus Kaplan-Meier are all biased and therefore inappropriate to quantify AE risk in the presence of varying follow-up times, CEs and censoring. Estimators are biased for estimation for absolute as well as relative AE risks. It is important to note that this bias is a statistical property of any of these estimators and independent of the purpose we use any of these estimators for, i.e., whether we quantify the risk for a prespecified or emerging AE, or estimate AE risk in a given therapeutic area, or want to detect a different AE signal between two treatment arms. Taking together, Stegherr et al.\cite{stegherr_20}, Stegherr et al.\cite{onesample}, and Rufibach et al.\cite{rufibach_23} provide theoretical (i.e. based on simulated) as well as empirical (based on data from 17 RCTs) evidence for the bias of all estimators apart from the gold-standard \aje, and also quantify this bias. This supports decade-long theoretical investigations into the bias of, e.g, the one minus Kaplan-Meier estimator in this field. We are of the strong opinion that all relevant stakeholders, among them clinicians, statisticians, regulators, should collaborate to finally implement fit-for-purpose methods to quantify AE risk, and update pertinent guidelines. In our opinion, the implementation of the ICH E9(R1) estimands addendum -- so far primarily for efficacy -- offers a window of opportunity to push for a change also in reporting safety information. In drug development, safety contributes as much to the benefit-risk assessment of a medicine as efficacy, so the same estimand, estimation, and reporting standards should apply to both.

\subsection{SAVVY -- template for sharing of summary data}

A special feature of SAVVY was they way data from the 17 RCTs was shared and analyzed: In a big collaborative effort data had been gathered within ten sponsor organisations (nine pharmaceutical companies and one academic trial center). In order to avoid challenges with data sharing SAVVY used an approach familiar from Health Informatics, see e.g. Budin et al.\cite{budin_15}. A standardized data structure was defined\cite{stegherr2019survival} based on which SAS and R macros were developed by the academic project group members. These macros where then shared with all participating sponsor organisations and run by them locally on their individual trial data. Only aggregated data necessary for meta-analyses were forwarded to the academic group members to centrally run meta-analyses. 

This appears to be an approach that could also be applied in other applications, as long as the analysis of interest can be done on summary data provided by single organisations.

\subsection{Roadmap for the SAVVY project and the analysis of safety data}

Work within the SAVVY project continues. Concrete plans are an analysis restricted to the oncology trials within the 17 RCTs, discussing in more detail the issue of CEs in a typical oncology clinical trial. Collaborations with clinicians in other therapeutic areas to define AEs of interest for which "proper" estimation of their risk would be informative and what clinical events to define as competing are envisaged. 

SAVVY's long-term vision is indeed to further familiarize trialists with the \aje and have this method be recommended in future revisions of pertinent guidelines. This, in connection with developing pragmatic approaches how to properly identify and define CEs in therapeutic areas.

\section{Abbreviations}

AE: adverse event; CE: competing event; SAVVY: Survival analysis for AdVers events with VarYing follow-up times; HR: hazard ratio, RR: relative risk; RCT: randomized clinical trial

\section{Declarations}

\section{Ethics approval and consent to participate}
Not applicable.

\section{Consent for publication}
Not applicable.

\section{Availability of data and materials}

Individual trial data analyses were run within the sponsor organizations using SAS and R software provided by the academic project group members. Only aggregated data necessary for meta-analyses were shared and meta-analyses were run centrally at the academic institutions. The SAVVY project has a webpage: \url{https://numbersman77.github.io/savvy}. Methods are implemented in the R package savvyr\cite{savvyr}. This package is available from CRAN.

\section{Competing interests}
KR is employee of F.\ Hoffmann-La Roche (Basel, Switzerland). JB has received personal fees for consultancy from Pfizer and Roche, all outside the submitted work. TF has received personal fees for consultancies (including data monitoring committees) from Bayer, Boehringer Ingelheim, Janssen, Novartis and Roche, all outside the submitted work. CS has received personal fees for consultancies (including data monitoring committees) from Novartis and Roche, all outside the submitted work. RS has declared no conflict of interest.

\section{Funding}
Not applicable.

\section{Author's contributions}
KR, JB, TF, CS and RS conceived the idea for article and drafted it. All authors critically reviewed the manuscript and approved its final version.

\section{Acknowledgements}

We thank Thomas K\"unzel and Daniel Saban\'{e}s Bov\'{e} for implementing earlier code into the R package savvyr\cite{savvyr}.


\bibliographystyle{ama}
\bibliography{savvy}

\end{document}